\documentclass[conference]{IEEEtran}
\usepackage{setspace}
\usepackage[english]{babel}
\usepackage[utf8]{inputenc}
\usepackage{enumerate}
\usepackage{graphicx}
\usepackage[square,sort,comma,numbers]{natbib}
\usepackage{multirow}
\usepackage{soul}
\usepackage{xcolor}
\usepackage[skip=0.5\baselineskip]{caption}
\usepackage{float}
\usepackage{caption}
\usepackage{mathtools}
\usepackage{booktabs}
\usepackage{multirow}
\usepackage{fancyhdr}

\fancypagestyle{firstpage}{%
    \lfoot{Prepint: under consideration at Pattern Recognition Letters}
}

\begin{document}
\title{On the performance of residual block design alternatives in convolutional neural networks for end-to-end audio classification}  

\author{\IEEEauthorblockN{Javier Naranjo-Alcazar\textsuperscript{1}, Sergi Perez-Castanos\textsuperscript{1}, Irene Martin-Morato\textsuperscript{2}, Pedro Zuccarello\textsuperscript{1}, Maximo Cobos\textsuperscript{2}} \\

\IEEEauthorblockA{\textsuperscript{1} \textit{Visualfy, Benisanó, Spain}}
\IEEEauthorblockA{\textsuperscript{2} \textit{Universitat de València, València, Spain}}\\

\IEEEauthorblockA{\textsuperscript{1} $\lbrace$\textit{javier.naranjo, sergi.perez, pedro.zuccarello}$\rbrace$\textit{@visualfy.com}}

\IEEEauthorblockA{\textsuperscript{2}$\lbrace$\textit{irene.martin, maximo.cobos$\rbrace$}\textit{@uv.es}}
}


\maketitle
\thispagestyle{firstpage}

\begin{abstract}
Residual learning is a recently proposed learning framework to facilitate the training of very deep neural networks. Residual blocks or units are made of a set of stacked layers, where the inputs are added back to their outputs with the aim of creating identity mappings. In practice, such identity mappings are accomplished by means of the so-called skip or residual connections. However, multiple implementation alternatives arise with respect to where such skip connections are applied within the set of stacked layers that make up a residual block. While ResNet architectures for image classification using convolutional neural networks (CNNs) have been widely discussed in the literature, few works have adopted ResNet architectures so far for 1D audio classification tasks. Thus, the suitability of different residual block designs for raw audio classification is partly unknown. The purpose of this paper is to compare, analyze and discuss the performance of several residual block implementations, the most commonly used in image classification problems,  within a state-of-the-art CNN-based architecture for end-to-end audio classification using raw audio waveforms. For comparison purposes, we also analyze the performance of the residual blocks under a similar 2D architecture using a conventional time-frequency audio representation as input. The results show that the achieved accuracy is considerably dependent, not only on the specific residual block implementation, but also on the selected input normalization.

\end{abstract}

\begin{IEEEkeywords}
Audio classification, Convolutional Neural Networks, Residual Learning, Sound Event Detection, UrbanSound8k
\end{IEEEkeywords}

\section{Introduction} \label{introduction}
\par Audio event classification (AEC) is the problem of categorizing an audio sequence into exclusive classes \cite{adavanne2017report, zhang2015robust, FOGGIA201522}. Basically, AEC is aimed at recognizing and understanding the acoustic environment based on sound information. This is usually treated as a supervised learning problem where a set of text-labels (such as siren, dog barking, etc.) describe the content of the different sound clips. In contrast to classical classification schemes based on feature extraction followed by classification, Deep Neural Networks (DNNs) \cite{lecun2015deep} reduce these steps by working as feature extractors and classifiers altogether. Amongst the many different deep learning techniques the ones based on Convolutional Neural Networks (CNNs) have shown very successful results in areas such as image classification or verification \cite{simonyan2014very, krizhevsky2012imagenet, szegedy2016rethinking, chollet2017xception}. CNNs are able to learn spatial or time invariant features from pixels (i.e. image) or from time-domain waveforms (i.e. audio signals). Several convolutional layers can be stacked to get different levels of representation of the input signal. Recently, CNNs have been proposed to treat audio related problems such as sound event detection or audio tagging, amongst many others \cite{zhang2019acoustic, hershey2017cnn, xu2018large}.

\par Although audio signals are natively one-dimensional sequences, most state-of-the-art approaches to audio classification based on CNNs use a two dimensional (2D) input \cite{cakir2016domestic, cakir2017convolutional}. Usually, these 2D inputs computed from the audio signal are well-known time-frequency representations such as Mel-spectrograms or the output of constant-Q transform (CTQ) filterbanks. Time-frequency 2D audio representations are able to accurately extract acoustically meaningful patterns but require a set of parameters to be specified, such as the window type and length or the hop size, which may have different optimal settings depending on the particular problem being treated or the particular type of input signals. In order to overcome these problems and providing an end-to-end solution, other approaches have proposed the use of 1D convolutions accepting the raw audio as input. Recent works show satisfactory results using these last kind of inputs and architectures \cite{dai2017very,qu2016understanding,lee2018samplecnn,gong2018deep,lee2017sample}. Note, however, that optimizing CNN hyperparameters is a challenging issue \cite{CUI2019}.

\par The present work is focused on the analysis of the performance of a particular CNN architecture, called Residual Network (ResNet), fed with 1D audio data. The ResNet architecture was first introduced in \cite{he2016deep} with the purpose of dealing with the vanishing gradient issue. The core idea of ResNet is to introduce the so-called \textit{identity weight shortcut connection} that skips one or more layers and adds the input of such layers to their stacked output. After the first residual unit was presented in \cite{he2016deep}, an exhaustive analysis of different variations of such a configuration was done for CNNs with 2D input signals to tackle the image classification problem \cite{he2016identity}. Nevertheless, although new residual blocks have been published in the context of 1D raw audio input waveforms \cite{lee2017sample}, no study has been carried out comparing different options for the residual blocks in the audio domain. The main objective of the present work is to analyze the performance and suitability of those residual blocks most commonly used for image classification problems, adapted to the context of 1D raw audio classification. To this end, six different residual blocks implementations that have shown good results in image classification problems are tested. Each of these blocks provide a varying scheme with regard to where identity mappings are created. All of them are analyzed under the common baseline architecture of \cite{dai2017very}, which presented a 1D CNN for raw audio waveform classification using the public urban-sound database UrbanSound8k  \footnote{https://urbansounddataset.weebly.com/urbansound8k.html}. For comparisson purposes, in this work, besides the UrbanSound8k, the public database ESC-50 \footnote{https://github.com/karoldvl/ESC-50} (concretely, the ESC-10 subset) has also been used for testing.

In addition, the performance of a 2D equivalent structure using a 2D time-frequency-based input representation is also provided. The experiments reveal that, while competitive results are obtained by such 1D ResNet architectures, the performance of a given residual block design is very dependent on the selected raw input normalization, which also motivates the use of 1D residual CNNs over 2D audio representations.

\begin{figure}[t]
\includegraphics[width=0.35\columnwidth]{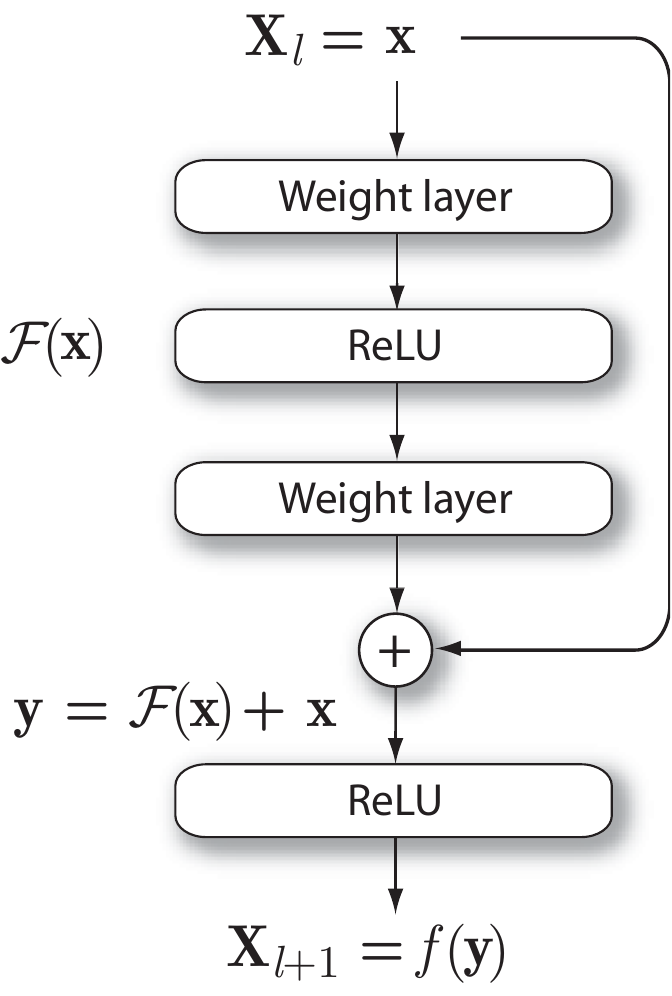}
\centering
\captionsetup{justification=centering}
\caption{Originally proposed residual block or unit \cite{he2016deep}.}
\label{fig:residual_unit}
\end{figure}

\begin{figure*}[t]
\includegraphics[width=1\textwidth]{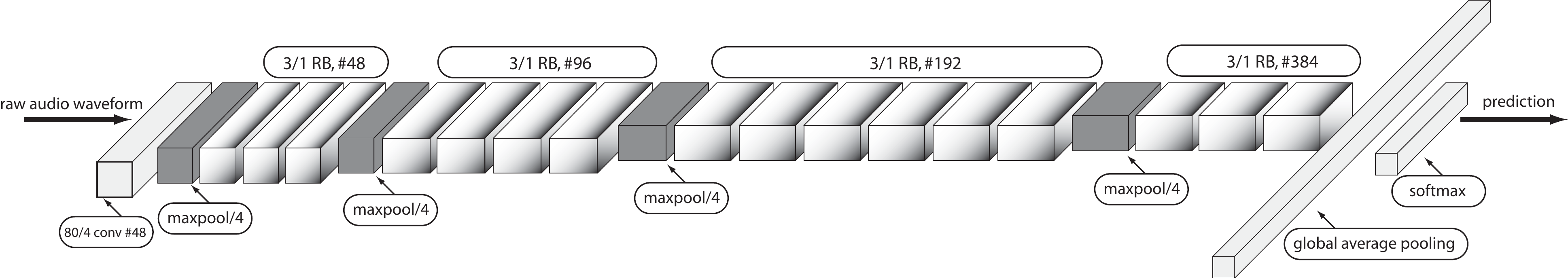}
\centering
\captionsetup{justification=centering}
\caption{Network analyzed \cite{dai2017very}. The architecture is explained as follows: [80/4, \#48]  denotes a layer with 48 filters, 80 of kernel size and stride equal to 4. RB blocks are indicated with kernel size, stride and number of filters. Each block of the diagram represents a layer.}
\label{fig:architecture_3d}
\end{figure*}

\section{Background}
\label{background}

The original residual block proposed in \cite{he2016deep} is shown in Fig.~\ref{fig:residual_unit}. 
Consider $\mathcal{H}(\mathbf{x})$ an underlying mapping to be fit by a set of stacked layers, where $\mathbf{x}$ is the input to the first of such layers. Residual blocks are designed to let such layers approximate a residual function, $\mathcal{F}(\mathbf{x}) \coloneqq \mathcal{H}(\mathbf{x})-\mathbf{x}$, which means that the original function can be expressed as $\mathcal{H}(\mathbf{x})~=~\mathcal{F}(\mathbf{x}) + \mathbf{x}$. The motivation of using residual blocks comes from the intuition that it may be easier to optimize the above residual mapping than to optimize
the original, unreferenced mapping. A straightforward way of implementing residual learning is by adding shortcut connections performing identity mapping. In such connections, the input to the set of layers $\mathbf{x}$ is added back to their output, so that $\mathbf{y} = \mathbf{x} + \mathcal{F}(\mathbf{x})$. The function $\mathcal{F}(\mathbf{x})$ represents the residual to be learned by a set of stacked layers of the CNN, where the weight layers are convolutional. In the original residual block, Rectified linear unit (ReLU) activation is applied to the result after each identity mapping, resulting in a final output $f(\mathbf{y})$ that acts as input to the next residual block, where $f(\cdot)$ denotes the ReLU function. Thus, in general, the input to the $l$-th block, $\mathbf{X}_l$, is the output from the previous block and its output becomes the input to the next one, $\mathbf{X}_{l+1}$. Note that shortcut connections do not add extra parameters nor additional computational cost. Thus, deeper networks can be trained with little additional effort, reducing substantially vanishing-gradient problems. Note, however, that CNNs often include Batch Normalization (BN) layers and vary in regards to where the activation function is applied. Therefore, the performance of residual learning may also depend both on the order followed by these layers and on the selected point at which shortcut connections are established. In \cite{he2016identity}, a careful discussion on identity mappings is provided, proposing the use of pre-activated residual units where $f$ is also an identity mapping, i.e. $\mathbf{X}_{l+1} = \mathbf{y}_l$. Such slight modification is shown to benefit the training process and to achieve better results in image recognition tasks. However, such analysis has only been performed for 2D architectures and, to the best of the authors' knowledge, a similar study analyzing residual blocks in 1D CNNs has not been addressed. The next section presents the residual block alternatives considered in this work.

\section{Network Architecture} \label{network}

\par All of the networks proposed in \cite{dai2017very}, labeled as M3, M5, M11, M18 and M34-res  in the original paper, share the same philosophy: they are fully-convolutional, intercalating convolutional and pooling layers. Fully-convolutional networks are, usually, able to obtain better generalization in the classified categories, whereas, fully-connected layers at the end of the network are more prone to show overfitting. In \cite{dai2017very}, the convolutional layers are configured with small receptive fields, with the exception of the first layer, whose receptive field is bigger in order to emulate a band-pass filter. Therefore, temporal resolution is reduced in the first two layers with large convolution and max pooling strides. After these layers, resolution reduction is complemented by doubling the number of filters in specific layers. Finally, after the last residual unit, global average pooling is applied to reduce each feature into a single value by averaging the activation across the input. To study the behavior of a given residual block (RB), this paper focuses on the M34-res architecture \cite{dai2017very} proposed for raw audio waveforms, which follows the general architecture shown in Fig. \ref{fig:architecture_3d}. 



\begin{figure*}[t]
\includegraphics[width=1\textwidth]{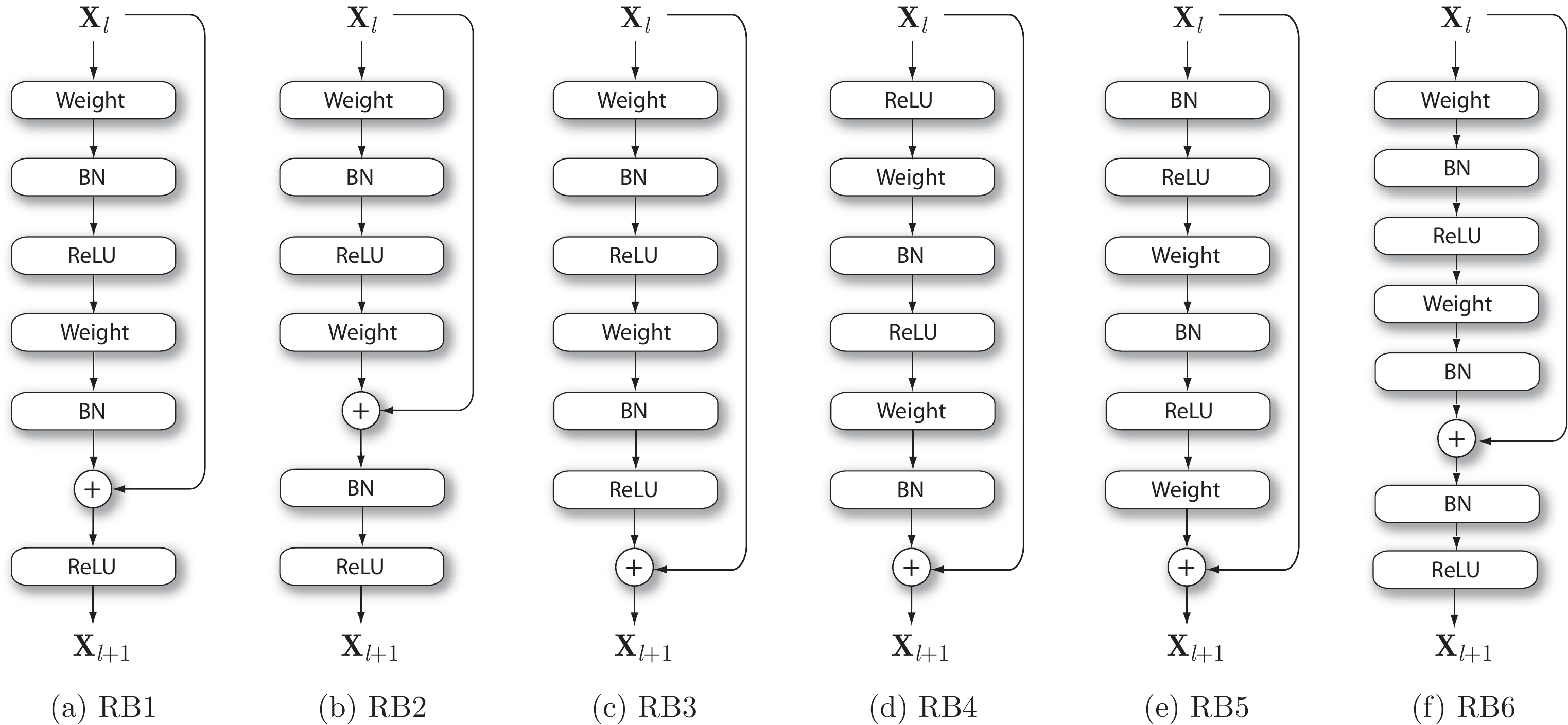}
\centering
\captionsetup{justification=centering}
\caption{Residual units implemented in this work. RB1 to RB5 (a-e) were first introduced in \cite{he2016identity}, whereas RB6 (f) was presented in \cite{dai2017very}.}
\label{fig:residual_5}
\end{figure*}

Six different RB implementation alternatives are analyzed: the original block proposed by He \textit{et al.} \cite{he2016deep} plus the other four blocks proposed by the same authors in \cite{he2016identity} and the one introduced by Dai \textit{et al.} in \cite{dai2017very} (see Fig.~\ref{fig:residual_5}). In the ResNets, the convolutional layers are replaced by the different RBs. To isolate the effect of these blocks from the rest of parameters of the network, the number of filters, the receptive field size and the number of convolutional layers remain the same as in \cite{dai2017very}. The analyzed residual blocks are the following:

\begin{itemize}
  \item \textbf{RB1 \cite{he2016deep}}: the input is first convolved and the output of the second convolution is the input of a batch normalization layer. After the addition, ReLU activation is applied.
  \item \textbf{RB2 \cite{he2016identity}}: the input is first convolved and no post-processing is done after the second convolution. The output of the addition is normalized and then activated using the ReLU function.
  \item \textbf{RB3 \cite{he2016identity}}: the input is first convolved as in \cite{dai2017very} and it the activation is performed before the addition.
  \item \textbf{RB4 \cite{he2016identity}}: the input is first passed through a ReLU activation layer and then normalized after the second convolution. There are no layers after the addition.
  \item \textbf{RB5 \cite{he2016identity}}: the input is first normalized and there are no layers after the second convolution as well as after the addition.
  \item \textbf{RB6 \cite{dai2017very}}: the input is first convolved and the output of the second convolution is the input of a batch normalization layer. After the addition, a new normalization is applied followed by ReLU activation.
\end{itemize}


The M34-res presented in \cite{dai2017very} has 4,001,242 parameters. RB5 has 3,988,570 parameters and the others have 3,989,914 parameters. Dropout layers have not been implemented neither after the pooling layers nor in the residual block.

\par Since ResNet architectures were originally proposed for image classification and many state-of-the-art audio classification systems are based on 2D time-frequency representations, we also tested the above RBs using a 2D ResNet-based CNN. The 2D architecture is a slim version of the previously described network, which accepts as input a log-Mel-spectrogram with 64 filters, using a window length of 40 ms with 50\% overlap. Frequency channels are scaled to zero mean and unit standard deviation, resulting in an input shape of (64 $\times$ number of time windows). When the audio input is shorter than the specified seconds depending on the dataset (see Subsection \ref{dataset}), the spectrogram is correspondingly padded with zeros on the time axis. As the network would have a huge number of parameters, some modifications have been applied to make the resulting network comparable to the 1D case. The first convolution and pooling layers are applied on the time axis by means of $1\times 80$ convolution and $1\times 4$ maxpooling. Note, however, that the rest of parameters are two-dimensional, as specified in Table \ref{tab:slim_m_34}. The number of repetitions of the RB blocks have been also modified to get a similar number of parameters, resulting in a 2D network with 4.16M parameters.

\begin{table}[]
    \centering
    \captionsetup{justification=centering}
    \begin{tabular}{cc}
        \hline
         &  [(3,3) RB, \#48] $\times 1$ \\
         \hline
         &  (4,4) max pooling \\
         \hline
         &  [(3,3) RB, \#96] $\times 2$ \\
         \hline
         &  (4,4) max pooling \\
         \hline
         &  [(3,3) RB, \#192] $\times 3$ \\
         \hline
         &  (4,4) max pooling \\
         \hline
         &  [(3,3) RB, \#384] $\times 1$ \\
         \hline
         &  (4,4) max pooling \\
         \hline
         & global average pooling \\
         \hline
         & softmax \\
         \hline
    \end{tabular}
    \caption{Slim version of the residual-based architecture for 2D audio representations.}
    \label{tab:slim_m_34}
\end{table} 

\section{Experimental Details} \label{experimental}

\subsection{Datasets and implementation details}\label{dataset}

\par As in \cite{dai2017very}, the experimental setup of the present work is based on UrbanSound8k \cite{salamon2014dataset}, a public sound-database that contains 8732 sound clips of duration of up to 4 seconds with 10 different classes such as dog barking, car horn, drilling, etc. The ESC-10 dataset \cite{piczak2015esc}, a public sound-database that contains 400 clips of 5 seconds of duration with 10 different categories (40 samples each category), is also considered. This dataset contains the same number of categories than UrbanSound8k, making the comparison more precise. When experimenting with UrbanSound8k, a 10-fold criteria was followed, 7895 sounds are used for training and 837 for validation, saving fold-10 for testing. Clips were resampled to 8~kHz and padded with zeros to reach 4~s length if necessary. Same procedure was run over ESC-10, in this case, as the dataset is partitioned in 5 folds, fold-5 was used for validation.

\begin{table*}[t!]
\centering
\caption{M34-res results with different normalization of the input signal and different residual blocks.}
\begin{tabular}{@{}cccccccc@{}}
\toprule
Pre-processing & Dataset & RB1 & RB2 & RB3 & RB4 & RB5 & RB6 \\ \midrule \midrule
\multirow{2}{*}{No} & UBS8k & 70.28$\pm$2.84\%& \textbf{72.07$\pm$1.56\%} & 66.62$\pm$5.05\% & 70.14$\pm$2.36\% & 71.31$\pm$1.63\% & \textit{71.52$\pm$1.96\%} \\
 & ESC-10 & \textbf{77.50$\pm$4.00\%} & 73.50$\pm$5.33\% & 73.25$\pm$7.71\% & 76.38$\pm$4.02\% & 59.38$\pm$7.46\% & \textit{77.00$\pm$8.23\%} \\ \midrule
\multirow{2}{*}{Scale max} & UBS8k & 73.11$\pm$1.74\% & \textbf{73.24$\pm$1.27\%} & 65.94$\pm$5.90\% & 72.89$\pm$1.58\% & 73.18$\pm$1.36\% & \textit{71.88$\pm$1.43\%} \\
 & ESC-10 & \textbf{78.50$\pm$5.52\%} & 75.75$\pm$6.62\% & 71.63$\pm$4.93\% & 75.50$\pm$7.53\% & 57.38$\pm$4.66\% & \textit{73.88$\pm$6.83\%} \\ \midrule
\multirow{2}{*}{Mean 0 Std 1} & UBS8k & 73.59$\pm$2.38\% & 72.98$\pm$0.96\% & 62.72$\pm$6.87\% & $\textbf{74.37$\pm$2.58\%}$ & 71.85$\pm$5.13\% & \textit{72.68$\pm$0.99\%} \\
 & ESC-10 & \textbf{81.38$\pm$4.95\%} & 80.50$\pm$4.61\% & 59.38$\pm$13.24\% & 73.38$\pm$9.05\% & 57.00$\pm$8.96\% & \textit{77.58$\pm$4.21\%} \\ \midrule
\multirow{2}{*}{log Mel-Spec} & UBS8k & 48.81$\pm$3.59\% & 66.94$\pm$2.05\% & 46.33$\pm$12.16\% & 50.03$\pm$9.05\% & 59.84$\pm$7.95\% & \textbf{\textit{67.67$\pm$1.35\%}} \\
 & ESC-10 & 51.63$\pm$14.19\% & \textbf{73.16$\pm$4.09\%} & 47.50$\pm$12.71\% & 50.38$\pm$11.70\% & 57.63$\pm$5.70\% & \textit{70.63$\pm$5.72\%} \\ \bottomrule
\end{tabular}
\label{tab:results}
\end{table*}

The optimizer used was Adam \cite{kingma2014adam}. The models were trained with a maximum of 400 epochs. Batch size was set to 128. The learning rate started with a value of 0.001 decreasing with a factor of 0.2 in case of no improvement in the validation accuracy after 15 epochs. The training is early stopped if the validation accuracy does not improve during 50 epochs. The initialization method was glorot-uniform and all weight parameters were subject to L2 regularization with a 0.0001 coefficient as in \cite{dai2017very}. Keras with Tensorflow backend were used to implement the models in the experiments. The audio manipulation module used in this work was LibROSA \cite{mcfee2015librosa}. Due to stochastic nature of the experiments each network was trained and tested 10 times. The mean and the standard deviation are reported in Table \ref{tab:results}. An observation with respect to the training of these networks is that the choice of the learning rate and its treatment has a considerable influence on the final result. This is reflected on the standard deviations of the configurations, suggesting that the networks may easily converge to different minima with slight changes in the parameters.

\section{Results} \label{results}

Results from our experiments are reported in Table \ref{tab:results}, where it can be observed that minor changes in the implementation of residual units across a baseline architecture have a considerable impact on the overall classification performance of the network. Moreover, the changes in performance are also very dependent on the selected input normalization. Interestingly, the best performances are obtained for different combinations of residual units and input normalizations, suggesting that different RB implementations may benefit better from different input pre-processings. According to the results shown in this table, RB1 and RB2 are usually the best performing ones in both datasets, but the general behavior of the RBs do not follow any identifiable pattern. The results of the 2D architecture using log-Mel-spectrograms as input representation are not as good as for the 1D architectures. In fact, it is also expected that different 2D audio representations will also affect considerably the final performance, complicating even more the design of the network.

\begin{figure}[t]
\includegraphics[width=1.0\columnwidth]{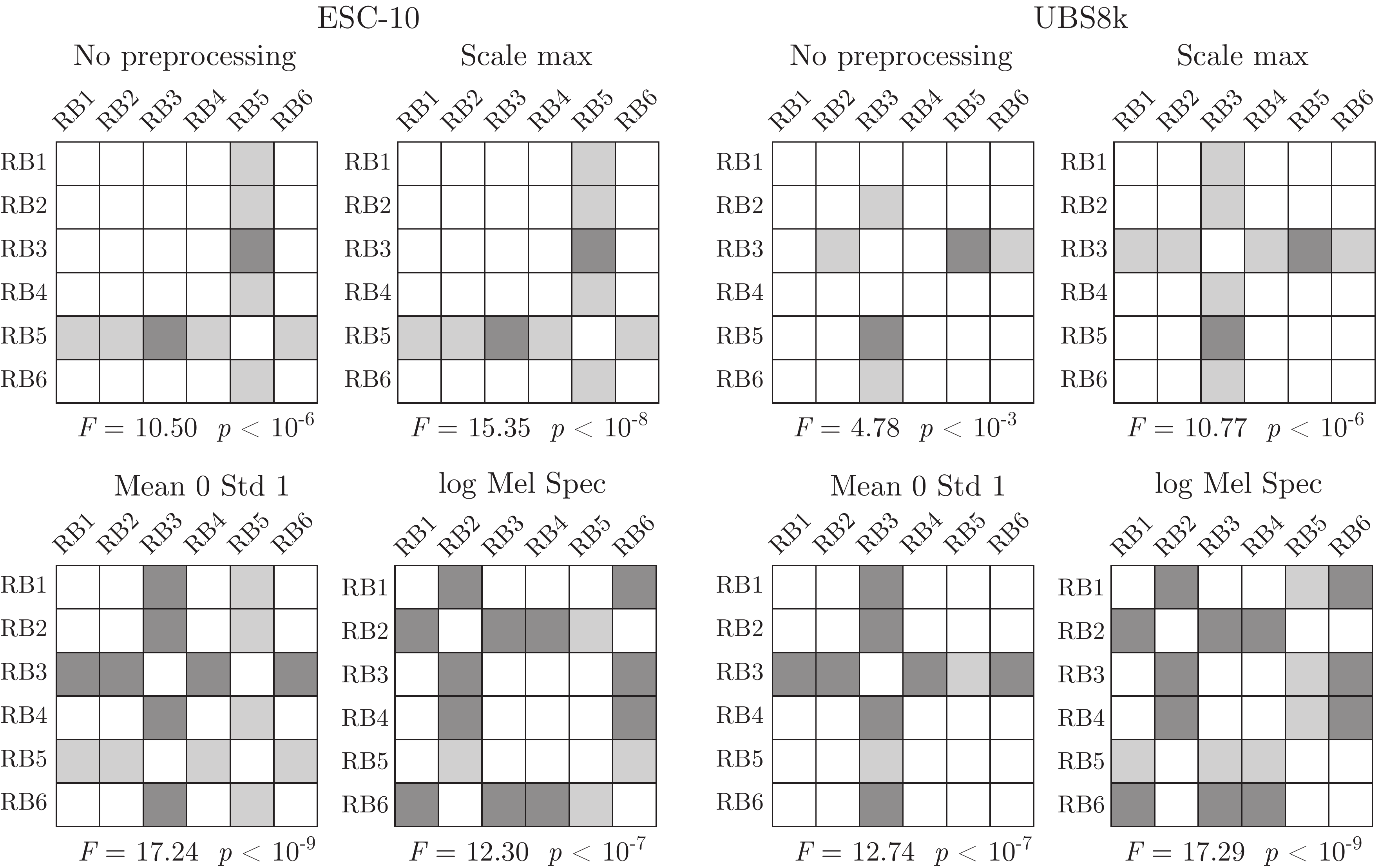}
\centering
\caption{Significance matrices for different pre-processings in each dataset. Shaded cells denote significant
differences as evaluated by the Tukey Kramer method. The darkest shade indicates matching significant differences in both datasets. $F$ and $p$ values for each one-way ANOVA are specified at the bottom of each matrix.}
\label{fig:significance}
\end{figure}

To provide further insight about these results, a multiple comparison based on the Tukey Kramer method \cite{Tukey1949}
has been carried out to examine the significant differences between the RBs for the selected pre-processings in both datasets. Figure~\ref{fig:significance} shows significance
matrices for each pre-processing, where a shaded cell denotes
a significant difference between the items in the corresponding
row and column. The corresponding $F$ and $p$ values obtained at each one-way ANOVA test are also specified at the bottom of each matrix. The darkest cells indicate a matching significant difference in both datasets for the same pre-processing. Although the differences in performance are, in general, quite dependent on the dataset and selected pre-processing, there are some matching differences across datasets. Specifically, the performances achieved by RB2 and RB5 are significantly different for ``no pre-processing" and ``max" normalization (Scale max). In contrast, when the normalization is applied in terms of mean and standard deviation, RB3 provides a performance significantly different from the rest. Finally, the differences observed in case of the 2D ResNet architecture show similar patterns in both datasets.

In order to compare with other classical approaches based on hand-crafted features and traditional classifiers, we take as a reference the values in \cite{piczak2015esc} for the ESC-10 dataset. Three baseline methods considering the means and standard deviations of 12 MFCCs and zero crossing rates were tested, achieving 66.7\% using k-NN, 72.7\% for random forest and 67.5\% for SVM. Similarly, the work by \cite{salamon2014dataset} achieved 70\% accuracy on the UrbanSound8k dataset using SVMs and random forests, using statistics (mean, variance, skewness, kurtosis, maximum, minimum, median) on 25 MFCCs and their first and second derivatives. The deep architectures analyzed in this paper can achieve better performance.

\section{Conclusion} \label{conclusion}

\par In this work, the performance of different residual units within a  1D CNN for end-to-end audio classification has been analyzed. While residual learning has been widely used in image recognition tasks, only few works have adopted ResNet architectures for audio-related tasks. With the objective of analyzing the RB alternatives, a baseline ResNet network \cite{dai2017very} for audio waveform classification has been selected. In addition, a similar 2D structure trained on audio log-Mel-Spectrograms has been also analyzed for comparison purposes. The results show that the performance achieved by a given RB is highly dependent on the selected input normalization, outperforming in some cases the baseline network. Moreover, the 1D networks achieved better results than the 2D ones, and tended to be easier to train. Thus, since the performance varies widely across RBs depending on the input, the design of ResNet architectures for 2D audio input representations could be significantly complex due to the variety of time-frequency transforms and the number of hyperparameters involved.





\section*{Acknowledgement}
This work was supported by the European Union Horizon 2020 research and innovation
programme [grant agreement No 779158]; and the Spanish Government [grant numbers RTI2018-097045-B-C21, PTQ-17-09106 and FPU14/06329].

\bibliographystyle{abbrv}
\bibliography{references}

\end{document}